\documentclass[sigplan,10pt,authorversion]{acmart}

\AtBeginDocument{%
  }

\acmYear{2025}\copyrightyear{2025}
\setcopyright{acmlicensed}
\acmConference[KISV '25]{3rd Workshop on Kernel Isolation, Safety and Verification}{October 13--16, 2025}{Seoul, Republic of Korea}
\acmBooktitle{3rd Workshop on Kernel Isolation, Safety and Verification (KISV '25), October 13--16, 2025, Seoul, Republic of Korea}
\acmDOI{10.1145/3765889.3767045}
\acmISBN{979-8-4007-2202-8/25/10}

\DeclareUnicodeCharacter{03BC}{\textmu}

\begin{document}

\title[Joyride: Rethinking Linux's network stack design]{Joyride: Rethinking Linux's network stack design for better performance, security, and reliability}

\author{Yanlin Du}
\orcid{0009-0004-9639-0871}
\affiliation{%
  \institution{The Pennsylvania State University}
  \city{University Park}
  \state{PA}
  \country{USA}
}
\email{duyanlin@psu.edu}

\author{Ruslan Nikolaev}
\orcid{0000-0002-1699-0593}
\affiliation{%
  \institution{The Pennsylvania State University}
  \city{University Park, PA}
  \state{PA}
  \country{USA}
}
\email{rnikola@psu.edu}

\begin{abstract}
Contemporary distributed computing workloads, including scientific computation, data mining, and machine learning, increasingly demand OS networking with minimal latency as well as high throughput, security, and reliability. However, Linux's conventional TCP/IP stack becomes increasingly problematic for high-end NICs, particularly those operating at 100 Gbps and beyond. 

These limitations come mainly from overheads associated with kernel space processing, mode switching, and data copying in the legacy architecture. Although kernel bypass techniques such as DPDK and RDMA offer alternatives, they introduce significant adoption barriers: both often require extensive application redesign, and RDMA is not universally available on commodity hardware.

This paper proposes Joyride, a high performance framework with a grand vision of replacing Linux's legacy network stack while providing compatibility with existing applications. Joyride aims to integrate kernel bypass ideas, specifically DPDK and a user-space TCP/IP stack, while designing a microkernel-style architecture for Linux networking.
\end{abstract}



\maketitle

\section{Introduction}

Modern distributed computing workloads, from scientific computation to machine learning, all rely on networks to parallelize their jobs to achieve better performance and accelerate computation. In these scenarios, network performance becomes their critical bottleneck. While network hardware has scaled from 10 Gbps to 100 Gbps and beyond, Linux's TCP/IP stack architecture has not evolved at the same pace. 

Recent kernel improvements like BIG TCP~\cite{BIGTCP} provide incremental benefits but still require substantial CPU resources to fully utilize available bandwidth. Our measurements show that Linux requires 4 to 8 CPU cores to saturate one 100 Gbps NIC. This inefficiency becomes increasingly problematic as networks evolve toward 200 Gbps and 400 Gbps.

This paper proposes Joyride, an architectural vision for Linux networking that would replace the kernel's TCP/IP stack with a transparent user-space implementation. Our key insight is that \emph{the problem is not the lack of high-performance networking solutions -- DPDK~\cite{dpdk}, RDMA, and various kernel bypass techniques that already exist, but rather that these solutions fragment the ecosystem by requiring application modifications, specific hardware, or exclusive NIC access.}
We propose a microkernel-inspired architecture where a centralized user-space network service would handle all TCP/IP processing while maintaining complete transparency to applications through modified LibC which intercepts and replaces network system calls.

The core challenge we seek to address is \emph{architectural}: How can we fundamentally restructure Linux networking for modern hardware without breaking the vast ecosystem of existing applications? Our plan is to combine three key design principles. First, we plan to leverage well-tested FreeBSD TCP/IP code running in user space rather than reimplementing TCP from scratch. That will ensure maximum reliability and protocol compliance. Second, we propose intercepting network system calls at the LibC level, which would enable transparent redirection without the need to modify or recompile applications. Third, we envision a system-wide deployment where \emph{all applications} share high-performance networking resources, unlike existing solutions that require per application configuration and exclusive hardware access.

Our evaluation on AMD EPYC 9005 series servers with 100 Gbps Intel E810 NICs shows that a single Linux process achieves less than 25 Gbps, while DPDK reaches near line rate with a single core. This 4× efficiency gap translates directly to increased costs and reduced performance.

Existing TCP user-space stacks have made important contributions but failed to provide a general-purpose replacement for Linux networking.

F-Stack~\cite{FSTACK} requires application modifications ranging from recompilation to complete restructuring around callback-based event handlers.
TAS~\cite{TAS} provides POSIX compatibility through library relinking but makes datacenter specific assumptions about network conditions (no IP fragmentation, reliable in order delivery, rare timeouts) that fail on the general Internet or with consumer hardware. 
Junction~\cite{junction} is optimized for cloud deployment density rather than general-purpose performance, with a focus on reducing memory footprint for containerized workloads.
libVMA~\cite{libvma} from NVIDIA/Mellanox provides LibC level translation but remains vendor specific. Pegasus~\cite{pegasus} advances transparent kernel bypass with support for both local and remote communication, addressing key challenges including LibC preloading and epoll/async APIs. Arrakis~\cite{arrakis}, while supporting POSIX interfaces, faces adoption challenges due to its custom OS architecture. RDMA based approaches demand specialized hardware on both endpoints, making them unsuitable for general Internet communication.
Finally, Gramine-TDX~\cite{gramine} offers another perspective on application compatibility with emerging network stacks, though focused on confidential computing rather than performance. Each solution excels in its target domain, but none provides the transparent, system-wide transformation needed for broad adoption in general-purpose systems~\cite{opentcp}.

This paper makes two primary contributions:
\begin{itemize}
\item We present a future design for a microkernel-inspired network stack architecture for Linux that decouples network processing from the kernel while maintaining full POSIX compatibility. Unlike prior work that targets controlled environments with predictable network conditions, our proposed design serves general-purpose Linux systems from laptops to servers, operating over arbitrary network conditions including the public Internet.

\item We quantify the growing performance gap between Linux's kernel network stack and modern NIC capabilities across multiple dimensions including throughput and CPU utilization. We demonstrate that this gap will only widen as network speeds increase, and analyze why recent kernel improvements provide only incremental gains rather than fundamental performance improvements.
\end{itemize}

We acknowledge that Joyride represents an early stage vision with a limited implementation. However, we provide the analysis of Linux's existing approach using state-of-the-art NICs. Our experimental results confirm that the architectural mismatch between Linux's kernel-based networking and modern hardware capabilities is a real and growing problem that requires fundamental rather than incremental solutions. While we have not yet implemented the full Joyride system, our measurements of the performance gap validate the urgent need for our proposed architecture.

\begin{figure}
  \centering
  \includegraphics[width=\columnwidth]{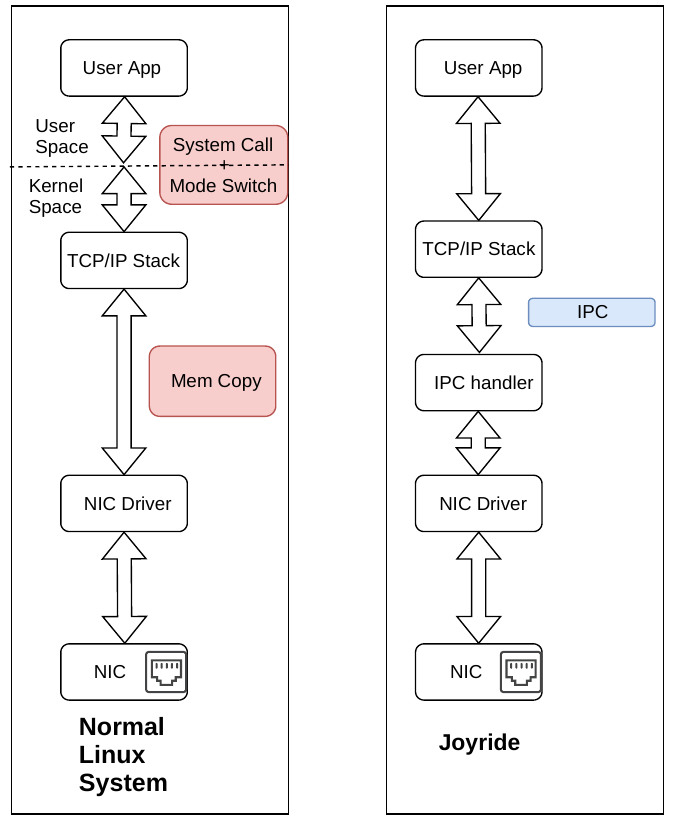}
  \caption{Traditional Linux network stack vs. Joyride.}
  \label{fig:comparison}
\end{figure}

\section{Background}
\subsection{Linux Network Stack Architecture}
The Linux kernel implements a complete TCP/IP stack that processes all network packets through multiple layers in kernel space. When an application sends data, it issues a system call that copies data from user space to kernel socket buffers. These buffers then traverse the TCP/IP stack and other network layers before reaching the NIC driver.

When a packet is received, the NIC triggers an interrupt (throttling techniques are used to avoid excessive interrupts), potentially causing a context switch to the kernel for packet processing. For high speed networks, this architecture imposes considerable overhead. A 100 Gbps link can deliver over 8 million packets per second with standard 1500 byte MTU, potentially triggering millions of interrupts and context switches. While optimizations like GRO (Generic Receive Offload) and GSO (Generic Segmentation Offload) reduce overhead, our measurements show that fundamental architectural costs persist, with 100 Gbps saturation still requiring at least 4 cores.

Recent kernel improvements have attempted to address these limitations. BIG TCP~\cite{BIGTCP} increases the maximum TCP segment size beyond 64KB, reducing per packet processing overhead. NetChannel~\cite{netchannel} introduces a disaggregated network stack architecture. While these optimizations provide measurable benefits, they operate within kernel constraints and cannot eliminate fundamental costs of context switching and data copying.

\subsection{DPDK and XDP}
Data Plane Development Kit (DPDK)~\cite{dpdk} is a high-performa\-nce packet processing framework that bypasses the kernel entirely to achieve line rate performance. It employs three key mechanisms: (1) poll mode drivers that continuously check for packets instead of using interrupts, thereby eliminating context switch overhead, (2) huge pages that reduce TLB misses and improve memory access efficiency, and (3) zero copy packet processing using ring buffers shared directly between user space and the NIC.

These optimizations enable DPDK to process packets at high rates with minimal CPU overhead. With 1500 byte packets, a single core can achieve 10 Gbps throughput, and with larger packets or optimized packet processing, even higher rates are possible. On modern systems with recent CPUs and NICs, DPDK can approach or reach 100 Gbps line rate with just one to two cores when using larger packet sizes. However, DPDK provides only raw packet I/O capabilities without implementing TCP/IP protocols, making it incompatible with existing socket based applications. This limitation has motivated several projects to build user-space TCP stacks on top of DPDK, including mTCP~\cite{mtcp}, F-Stack~\cite{FSTACK}, TAS~\cite{TAS}, and Junction~\cite{junction}, each attempting to bridge the gap between DPDK's performance and application compatibility with varying degrees of success.
Express Data Path (XDP)~\cite{xdp} provides an alternative approach that operates within the kernel but prior to the network stack. XDP programs run as eBPF code directly in the driver's receive path, enabling early packet filtering and forwarding decisions. While XDP offers lower latency than traditional kernel processing, it remains limited to simple packet operations. XDP could potentially serve as an underlying mechanism for some Joyride components, particularly for packet filtering and steering to user-space processors.

\subsection{User Space TCP}
Building TCP/IP stacks in user space introduces fundamental complexity beyond raw packet processing. The core TCP specifications (RFC 9293~\cite{RFC9293} and RFC 1122~\cite{RFC1122}) define only baseline protocol behavior. Production implementations must support over 20 extension RFCs covering selective acknowledgments, window scaling, timestamps, and congestion control algorithms beyond the core specifications.

Several projects have attempted user-space TCP stacks. IX~\cite{IX} provides a data plane OS using hardware virtualization. mTCP~\cite{mtcp} achieves multicore scalability through per-core instances. TAS~\cite{TAS} maintains socket compatibility through library interposition. Junction~\cite{junction} and libVMA~\cite{libvma} use LibC preloading for transparent socket interposition. Pegasus~\cite{pegasus} addresses transparent kernel bypass including epoll/async API support. F-Stack~\cite{FSTACK} ports FreeBSD's TCP stack into user space. Arrakis~\cite{arrakis} provides a custom OS architecture.

\subsection{SR-IOV (Single Root I/O Virtualization)}
SR-IOV~\cite{ms_sriov} is a PCIe standard that enables a single physical NIC to be partitioned into multiple virtual functions (VFs), each acting as an independent network device. This hardware level isolation makes SR-IOV widely deployed in virtualization.

SR-IOV enables practical coexistence between kernel and user-space networking on the same physical hardware. By partitioning a NIC into multiple VFs, system administrators can assign some VFs to the traditional Linux kernel stack while dedicating others to high-performance user-space stacks like DPDK. However, current deployments require manual configuration and explicit application binding to specific VFs. Joyride proposes to automate this process, transparently directing applications to appropriate network paths based on their performance requirements and system resources.

\subsection{Remote Direct Memory Access (RDMA)}
RDMA presents an alternative approach to high-performance networking that bypasses both the kernel and TCP/IP stack entirely. By enabling direct memory transfers between applications across hosts, RDMA eliminates kernel involvement, interrupts, and buffer copying, achieving microsecond latencies and near zero CPU overhead. Data moves directly between application memory and NIC hardware without any software processing in the data path.
However, RDMA's deployment requirements severely limit its applicability for general purpose networking. Both endpoints must have RDMA capable NICs. Besides that, applications must be explicitly written for RDMA APIs. These requirements are practical only in specialized environments like HPC clusters or private datacenters with homogeneous hardware. Furthermore, RDMA protocols struggle with lossy networks and cannot traverse standard Internet infrastructure without specialized gateways. These constraints make RDMA inappropriate as a general replacement for TCP/IP, despite its good performance characteristics in controlled environments.
The limitations of RDMA underscore why general purpose systems need solutions like Joyride that can improve performance while maintaining TCP/IP compatibility. Rather than bypassing TCP/IP entirely as RDMA does, our proposed architecture moves TCP/IP processing to user space while preserving standard socket semantics, enabling deployment across commodity hardware and existing networks without requiring specialized NICs or protocol support at both endpoints.

\section{Design}
\subsection{Limitations of Related Work}
While kernel bypass solutions demonstrate the potential for user-space networking, they fail to provide practical system-wide solutions. F-Stack~\cite{FSTACK}'s per-core process architecture creates significant deployment challenges: applications must be redesigned to use explicit shared memory for cross core communication, and event handlers must be restructured as callbacks that yield control back to the network stack. Its architectural constraints limit it to specific workloads.

TAS~\cite{TAS} addresses some architectural issues but makes datacenter specific assumptions that limit general applicability. The system assumes no IP fragmentation, reliable in-order packet delivery, and rare timeout events conditions that hold within controlled data centers but are unrealistic on the general Internet or with consumer hardware.

IX~\cite{IX} implements a data plane operating system that separates control plane from data plane using hardware virtualization. While IX achieves impressive performance, its design philosophy prioritizes isolation and protection over flexibility. The strict separation between control and data planes means applications must be entirely rewritten for the IX data plane API with batched system calls, unlike approaches that intercept standard socket calls. IX's run to completion model prohibits any blocking operations, requiring careful application design to avoid stalling the entire data plane. Most critically, IX lacks support for standard TCP extensions and compatibility features accumulated over decades of Internet evolution. The system implements a simplified TCP focused on datacenter scenarios, missing functionality that production deployments require. This makes IX suitable primarily for private data centers rather than as a replacement for existing systems.

Junction~\cite{junction} targets cloud environments specifically, optimizing for application density and memory efficiency rather than general purpose performance. Junction, in its fastest mode, supports only certain Mellanox NICs, which further limits its generality. While Junction achieves transparent socket interposition through binary rewriting, unlike our proposed LibC interception approach, it similarly lacks the system-wide deployment model we envision.

libVMA~\cite{libvma} provides mature LibC level translation but remains vendor specific to NVIDIA/Mellanox hardware. Pegasus~\cite{pegasus} represents recent advances in transparent kernel bypass, addressing many challenges including LibC preloading and epoll/async API support, but still requires per application configuration.

Arrakis~\cite{arrakis}, while providing POSIX interface support, requires dedicated CPU cores for its control plane and uses a custom OS architecture that diverges from standard Linux deployments. 

The fragmentation of these solutions, each excelling in its target domain but constrained by unique assumptions and requirements, underscores the need for a unified approach that provides transparent, system-wide high-performance networking across diverse hardware and network conditions while maintaining the flexibility of kernel networking.

\begin{figure}
  \centering
  \includegraphics[width=\columnwidth]{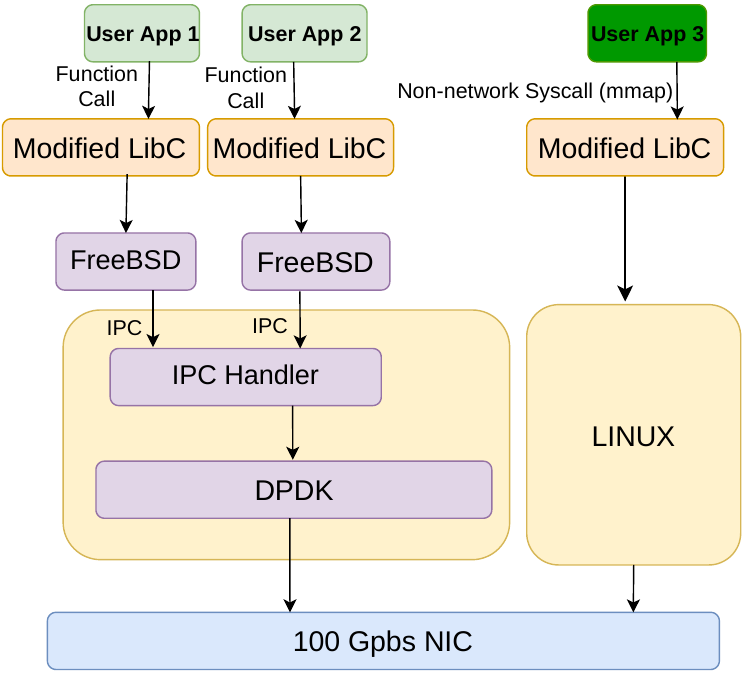}
  \caption{Proposed Joyride's architecture.}
  \label{fig:architecture}
\end{figure}

\subsection{Proposed Joyride's Architecture}
Figure~\ref{fig:comparison} compares the traditional Linux network stack with Joyride's architecture. Unlike existing approaches that require per application modifications or exclusive hardware access, we propose implementing a system-wide network stack replacement that would operate entirely in user space while maintaining complete transparency to applications.

The key insight driving our design is centralizing network processing in a dedicated network service process that would manage the physical NIC through DPDK. Applications would interact with this service through a modified LibC that intercepts all network related system calls (socket, send, recv, select, epoll, etc.) and redirects them to our user-space implementation. Non-network system calls would pass through unchanged to the kernel, maintaining full OS compatibility. This approach differs fundamentally from TAS, which requires explicit relinking, and from F-Stack, which demands application restructuring.

Figure~\ref{fig:architecture} presents our proposed architecture in detail. User applications (App 1 and App 2) requiring high-performance networking would use a modified LibC that transparently redirects network calls to our user-space stack. This stack would consist of FreeBSD derived TCP/IP instances coordinated through an IPC Handler built on DPDK. App 3 demonstrates the planned fallback path, where unmodified applications would transparently use the kernel stack. The network service would multiplex the single physical NIC among all applications while maintaining standard Linux network semantics, and applications would share the same IP address and network namespace. All packet processing would occur entirely in user space, eliminating the context switches between user and kernel space that limit traditional Linux networking performance.

Our design philosophy differs from existing solutions in three key aspects. First, we propose system-wide deployment rather than per application configuration. Second, we aim for complete transparency without requiring any application modifications or relinking. Third, we plan to support general Internet conditions including IP fragmentation and packet reordering, which datacenter-focused solutions like TAS optimize away.
Additionally, this microkernel-inspired design enhances security through privilege separation: while user-space services are still exposed to attacks, any compromise remains contained at user privilege level rather than gaining kernel access. At the same time, the reduced kernel image size decreases the trusted computing base.

\subsection{Security Considerations and Threat Model}
Joyride's centralized network service raises important security questions. Our threat model assumes that applications may be compromised, but the network service itself remains trusted (similar to trusting the kernel network stack).

When SR-IOV is available, we can support multiple isolated network stacks by using hardware-enforced VF isolation. Mission-critical applications would get full isolation from other applications by using separate instances of the network stack, or even running all network-stack components directly in their respective address spaces while using dedicated VFs. This is a hybrid model, which is similar to~\cite{10.1145/3381052.3381316}.

Without SR-IOV, isolation is fully software-based through memory protection and capability-based access control. Each application receives a dedicated shared memory region that other applications cannot access.
The IPC interface between applications and the network service uses capability-based authentication, where each application receives unforgeable tokens for its sockets. This prevents compromised applications from accessing other connections' data. While the centralized service could be seen as a single point of failure, it is no different from trusting the kernel, and the service can be restarted without system reboot if necessary.

\subsection{Transparent Integration Through LibC Interception}
We propose achieving zero-modification deployment through careful LibC integration. Our design would override network system calls at the dynamic-linking layer, redirecting them to our user-space stack while preserving exact POSIX semantics. This approach aims to succeed where F-Stack fails by implementing the complete network API surface, including complex operations like select/epoll multiplexing, non-blocking I/O, and signal handling.
The planned communication mechanism between applications and the network service would use shared memory for data transfer and lightweight signaling for control operations. Non-networking system calls would continue to be directed to the kernel by LibC, ensuring full compatibility.

Unlike TAS's library relinking approach, which requires explicit application or deployment configuration, our LibC interception would operate transparently through standard dynamic linking mechanisms. Applications would not need recompilation, relinking, or even awareness that they are using a user-space network stack. This transparency extends to complex applications using multiple libraries, as our interception would occur at the lowest level of the system call interface.

\subsection{Fallback and Protocol Completeness}
A critical aspect of our proposed design is combining robustness with performance through two mechanisms: automatic fallback and protocol completeness. If desired, Joyride can still leave a possibility for certain applications to fall back to the kernel network stack: we just need to load such applications with a different version of LibC and assign one VF to the kernel. This can be valuable and less disruptive if some applications rely on specific or cryptic kernel functionality, while most other applications would benefit from high-performance user-space networking. This flexibility provides best available performance while guaranteeing functionality, making it suitable for production environments where reliability matters more than peak performance.

For protocol implementation, we plan to use FreeBSD's mature TCP/IP stack as our foundation, ensuring correctness across the full complexity of real world networking. This includes over 20 RFC extensions beyond basic TCP, congestion control variants, and undocumented but necessary compatibility behaviors. Unlike TAS's datacenter optimizations that assume no fragmentation and reliable delivery -- assumptions that would cause failures on the general Internet -- our design would handle arbitrary network conditions, from perfect datacenter networks to lossy wireless connections. Rather than reimplementing TCP from scratch, we focus our efforts on the architectural challenges of integrating a user-space network stack transparently into Linux systems.
We are also exploring XDP~\cite{xdp} as a potential mechanism for packet steering and early filtering. XDP programs could redirect packets to user space via AF\_XDP sockets, providing a kernel sanctioned path for packet extraction that does not require exclusive NIC ownership.

\section{Preliminary Evaluation}
Our preliminary evaluation explores the system-level inefficiencies that prevent applications from achieving high-performance networking without sacrificing transparency or compatibility. We investigate the performance gap between kernel networking and user-space packet processing to understand why developers must choose between performance and compatibility.

We conduct our experiments on AMD EPYC 9005 series servers equipped with Intel E810 100 Gbps NICs running Ubuntu 22.04 with Linux kernel v6.2. We compare three configurations: (1) standard blocking sockets using Linux's kernel TCP/IP stack, (2) non-blocking sockets with the kernel stack to explore the impact of blocking overhead, and (3) raw DPDK packet processing to establish hardware capability limits. To ensure fair comparison with potential user-space implementations, we reimplemented \verb|ttcp|~\cite{ttcp} using non-blocking I/O with epoll, as some user-space TCP stacks (e.g., F-Stack) inherently operate in non-blocking mode due to the absence of kernel-assisted blocking primitives.

Traditional blocking sockets rely on kernel mechanisms to suspend and resume processes, incurring context switch overheads with each blocking operation. User-space network stacks cannot provide true blocking semantics without kernel assistance. Our non-blocking implementation of \verb|ttcp| helps isolate the performance impact of context switches from other kernel overheads. With standard blocking sockets, a single \verb|ttcp| process achieves less than 20 Gbps on our 100 Gbps link. Our non-blocking implementation improves this to approximately 25 Gbps, demonstrating that blocking overheads contribute to but do not fully explain the performance gap.

In contrast, raw DPDK packet processing reaches line rate (near 100 Gbps) with a single CPU core. This 4× performance gap between non-blocking kernel sockets and DPDK stems from fundamental architectural issues: system call overhead, kernel memory management, and lock contention. Figure 3 shows throughput scaling with different buffer sizes for both blocking and non-blocking implementations. The non-blocking variant generally outperforms blocking sockets, though we observe unexpected behavior at 8KB buffer size that requires further investigation. Both implementations plateau far below link capacity.

Figure 4 illustrates aggregate throughput when scaling processes or cores. The kernel network stack requires 4 to 8 cores to approach link saturation. This reveals another limitation: kernel lock contention and cache coherence overhead increase with core count. Our proposed user-space architecture would reduce kernel synchronization bottlenecks, enabling each core to process packets independently without competing for shared kernel resources.

We acknowledge that our current evaluation compares kernel-based TCP/IP against raw DPDK packet processing rather than a complete user-space TCP implementation. However, the 4× gap is substantial and demonstrates the potential benefits of bypassing the kernel, though actual user-space TCP performance would be lower than raw DPDK due to protocol processing overheads.

\begin{figure}
  \centering
  \begin{minipage}{0.49\columnwidth}
    \centering
    \includegraphics[width=\textwidth]{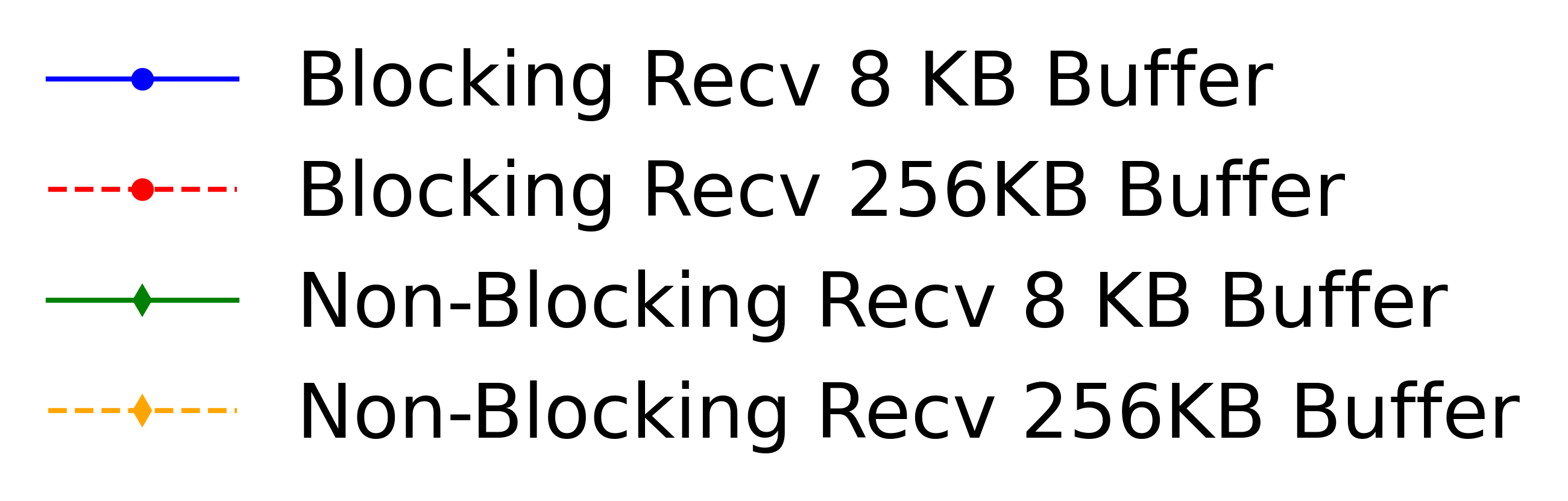}
  \end{minipage}
  \hfill
  \begin{minipage}{0.49\columnwidth}
    \centering
    \includegraphics[width=\textwidth]{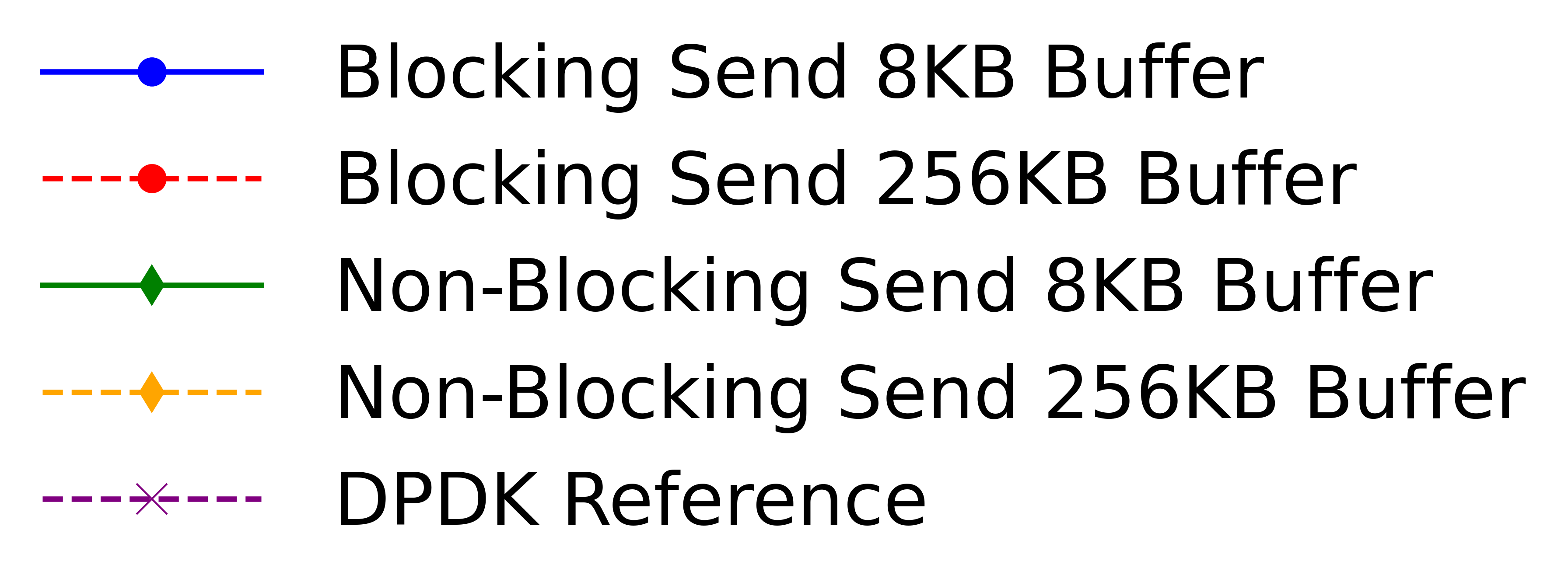}
  \end{minipage}
  \\[0.2cm]
  \begin{minipage}{0.49\columnwidth}
    \centering
    \includegraphics[width=\textwidth]{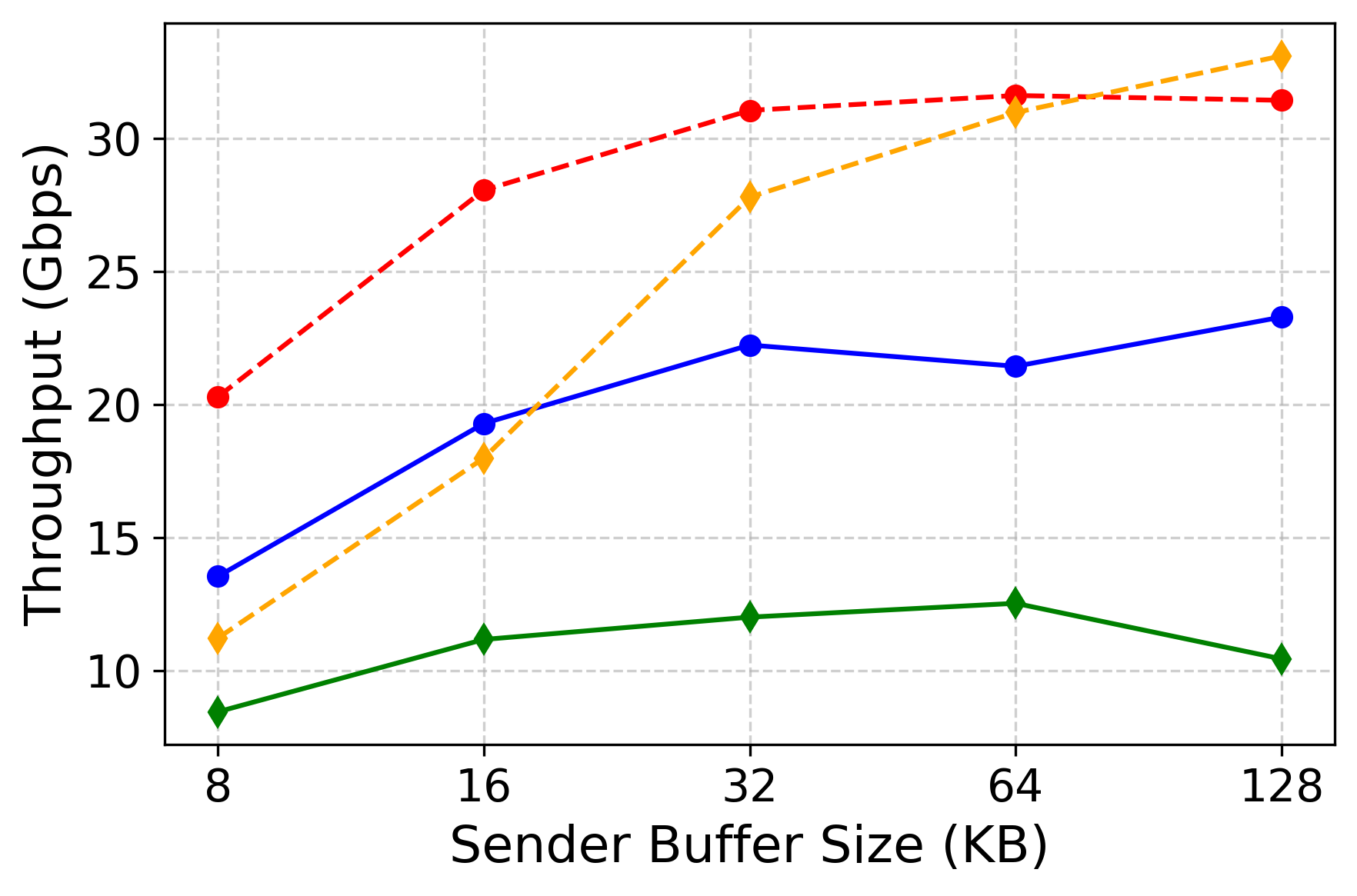}
    \caption{Throughput of a single process for different buffer sizes (Linux).}
    \label{fig:buffer_size}
  \end{minipage}
  \hfill
  \begin{minipage}{0.49\columnwidth}
    \centering
    \includegraphics[width=\textwidth]{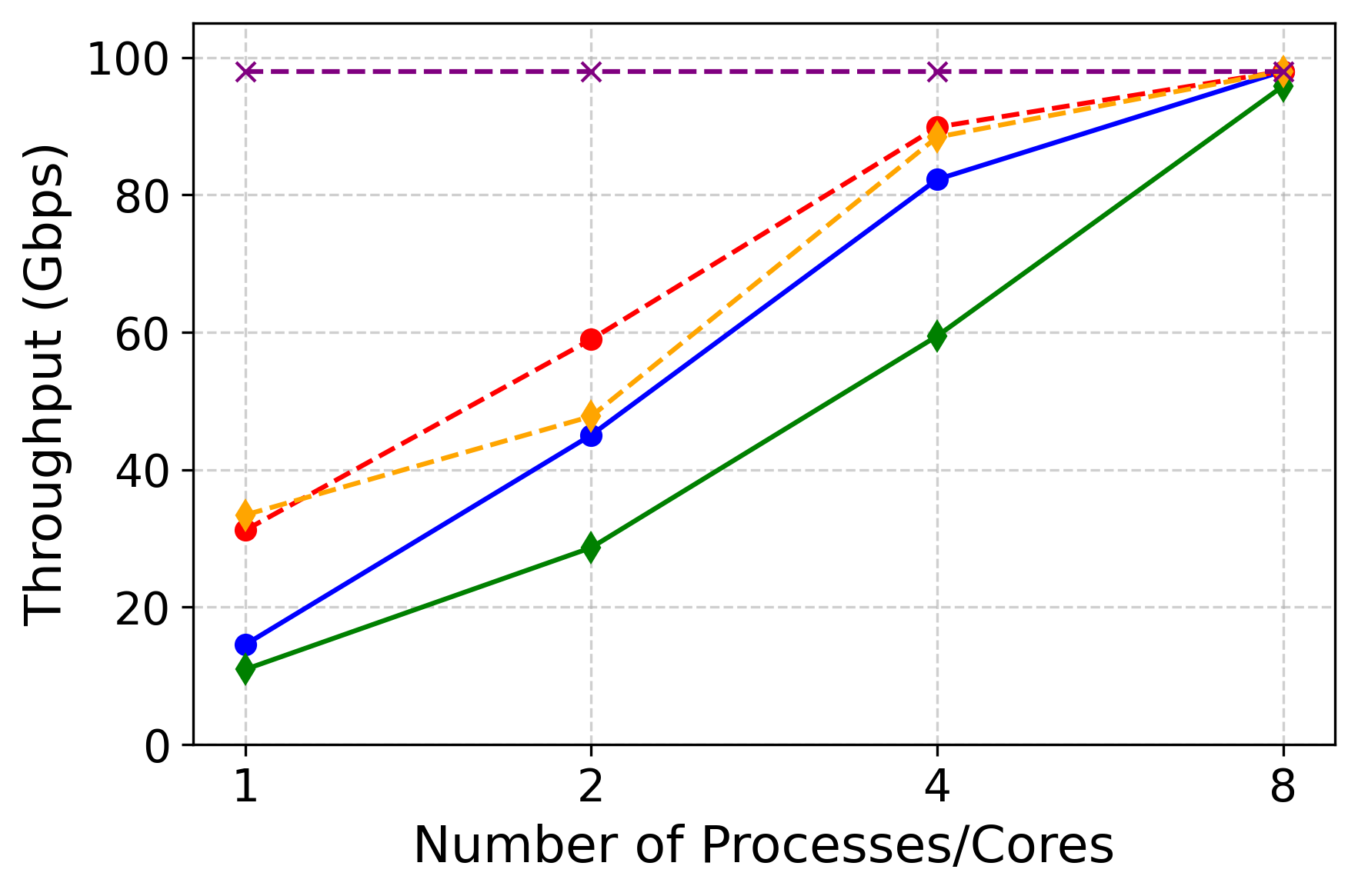}
    \caption{Aggregate network throughput (DPDK vs. Linux).}
    \label{fig:parallel_streams}
  \end{minipage}
  
  \label{fig:evaluation}
\end{figure}

\section{Conclusion and Future Work}
This paper presents Joyride, a proposed user-space networking framework that aims to replace Linux's traditional kernel-mode network stack while maintaining full application compatibility. Our evaluation demonstrates that Linux networking requires 4-8 cores to saturate a 100 Gbps NIC, while DPDK achieves line rate with a single core. As network speeds scale beyond 100 Gbps, this inefficiency becomes increasingly problematic.

Our architecture proposes to address kernel processing limitations through centralized user-space networking that enables transparent application support via LibC interception. Unlike existing solutions such as F-Stack, which requires callback restructuring, TAS, which needs relinking, or RDMA, which demands special hardware, Joyride targets seamless deployment for unmodified applications. The key insight is that transparency and performance need not be mutually exclusive if the architecture properly separates concerns between packet processing and application interfaces.

We are developing the DPDK-based translation layer and FreeBSD's TCP stack integration to realize this vision. Future work includes implementing full TCP and UDP stacks, developing the microkernel-inspired server based on DPDK, creating automated policies for dynamic fallback between user-space and kernel paths, and exploring XDP integration for packet steering. While significant implementation work remains, our preliminary results validate both the severity of current limitations and the potential for transparent user-space networking.

\bibliographystyle{ACM-Reference-Format}
\bibliography{references}

\end{document}